\documentclass[%
superscriptaddress,
reprint,
amsmath,amssymb,
aps,
prb,
]{revtex4-1}
\usepackage{graphicx}
\usepackage{dcolumn}
\usepackage{bm}
\usepackage{hyperref}

\begin{document}

\title{Realization of complex conjugate media using the real spectra of non-PT-symmetric photonic crystals}

\author{Xiaohan Cui}
\affiliation{%
	Department of Physics, The Hong Kong University of Science and Technology, Clear Water Bay, Kowloon, Hong Kong SAR, China
}%
\author{Kun Ding}
\affiliation{%
	Department of Physics, The Hong Kong University of Science and Technology, Clear Water Bay, Kowloon, Hong Kong SAR, China
}%

\affiliation{The Blackett Laboratory, Department of Physics, Imperial College London, London SW7 2AZ, United Kingdom}
\author{Jian-Wen Dong}
\affiliation{
School of Physics $ \& $ State Key Laboratory of Optoelectronic Materials and Technologies, Sun Yat-Sen University, Guangzhou 510275, China}
\author{C. T. Chan}%
 \email{phchan@ust.hk}
\affiliation{%
Department of Physics, The Hong Kong University of Science and Technology, Clear Water Bay, Kowloon, Hong Kong SAR, China
}%



\date{\today}
\begin{abstract}
 While parity-time (PT)-symmetric systems can exhibit real spectra in the exact PT-symmetry regime, the PT-symmetry is actually not a necessary condition for the real spectra. Here we show that non-PT-symmetric photonic crystals carrying Dirac-like cone dispersions can always exhibit real spectra as long as the average non-Hermiticity strength within the unit cell for the eigenstates is zero. By building a non-Hermitian Hamiltonian model, we find that the real spectra of the non-PT-symmetric system can be explained using the concept of pseudo-Hermiticity. We demonstrate using effective medium theories that in the long-wavelength limit, such non-PT-symmetric photonic crystals behave like the so-called complex conjugate medium (CCM) whose refractive index is real but whose permittivity and permeability are complex numbers. The real refractive index for this effective CCM is guaranteed by the real spectrum of the photonic crystals and the complex permittivity and permeability comes from the non-PT-symmetric loss-gain distributions. We show some interesting phenomena associated with the CCM, such as the lasing effect.
\end{abstract}
\maketitle

\section{Introduction} 
Hermitian Hamiltonian can describe ideal closed physical systems, in which the total energy is conserved, and eigenfrequencies are purely real. However, non-conservative elements are ubiquitous in classical wave systems, and we need to introduce the concept of non-Hermiticity to describe such systems. In the past decades, there is a surge of interest in studying the physics of non-Hermitian systems \cite{bender_real_1998,bender_complex_2002,mostafazadeh_exact_2003,bender_making_2007,rotter_non-hermitian_2009,mostafazadeh_pseudo-hermitian_2010,moiseyev_non-hermitian_2011} and the parity-time (PT)-symmetric system, which was first introduced in quantum mechanics \cite{bender_real_1998}, is the most popular one. A Hamiltonian $ H $ is said to be PT-symmetric if  $ [H,PT]=0 $, where the operator P represents a space reflection, and the operator T represents a time reversal. PT-symmetric systems have been realized in optics and photonics \cite{makris_beam_2008,ruter_observation_2010,regensburger_paritytime_2012}. In an optical system, the loss-gain (represented by the imaginary part of the refractive index) distribution is PT-symmetric, if the complex-valued refractive index satisfies $ n(x)=n^*(-x) $.

The most intriguing property of a PT-symmetric Hamiltonian is that it can exhibit real spectra in the exact PT-symmetry regime \cite{bender_complex_2002}. As we change the parameters of the Hamiltonian, such as the loss-gain strength, it may go into the broken PT-symmetry regime, and the eigenvalues form complex conjugate pairs. The symmetry-breaking point, marking the phase transition in the eigenvalue spectrum, is the exceptional point (EP) \cite{heiss_physics_2012,ding_emergence_2016}. 
However, PT-symmetry is not a necessary condition for achieving a real spectrum \cite{mostafazadeh_pseudo-hermiticity_2002-2} and some studies about realizing real spectra in non-PT-symmetric systems have been presented recently \cite{cannata_schrodinger_1998,tsoy_stable_2014,makris_constant-intensity_2015,nixon_all-real_2016,hang_realization_2017}. It has been proved that the necessary condition for the real spectrum is pseudo-Hermiticity \cite{mostafazadeh_pseudo-hermiticity_2002-2}. 
A Hamiltonian $ H $ satisfying $\eta H{\eta ^{ - 1}} = {H^\dag }$ , where $ \eta $ is an invertible linear Hermitian operator, is defined as pseudo-Hermitian Hamiltonian (see Supplementary Note 1 for details). 
Real spectra realized in non-PT-symmetric systems can be explained using the concept of pseudo-Hermiticity.

Achieving the real spectra in a non-PT-symmetric PC is highly desirable because from an effective medium theory (EMT) point of view \cite{choy_effective_2015}, realizing the real spectra in certain regions of the Brillouin zone can lead to new applications. For example, one can then realize a complex conjugate medium (CCM) \cite{dragoman_complex_2011,imran_effects_2011,basiri_light_2015}, which carries many unusual phenomena, including coherent perfect absorption and lasing \cite{bai_simultaneous_2016}, and negative refraction \cite{xu_electromagnetic_2017}. The refractive index of the CCM is a real number, but their permittivity and permeability can both be complex numbers in general \cite{dragoman_complex_2011}. From the EMT point of view, achieving a complex conjugate medium using a PC requires the non-Hermitian PC to exhibit real frequency bands near the Brillouin zone center ($ \Gamma $ point) because the effective refractive index is real. 

To understand the underlying physics, we build a two-band non-Hermitian Hamiltonian model for the non-PT-symmetric photonic crystals. We show that the model Hamiltonian is pseudo-Hermitian as long as the average non-Hermiticity strength (to be defined below) within the unit cell for the relevant states is zero and this condition can always be achieved in PCs that have Dirac-like cone dispersions. Consequently, real spectra can be realized in a particular frequency range, implying that the effective refractive index is real. In the long-wavelength limit, the scattering properties of such an inhomogeneous PC behave indeed like a homogeneous complex conjugate medium.

\section{The two-band non-Hermitian Hamiltonian model} 

\begin{figure}[htpb]
	\includegraphics[width=\linewidth]{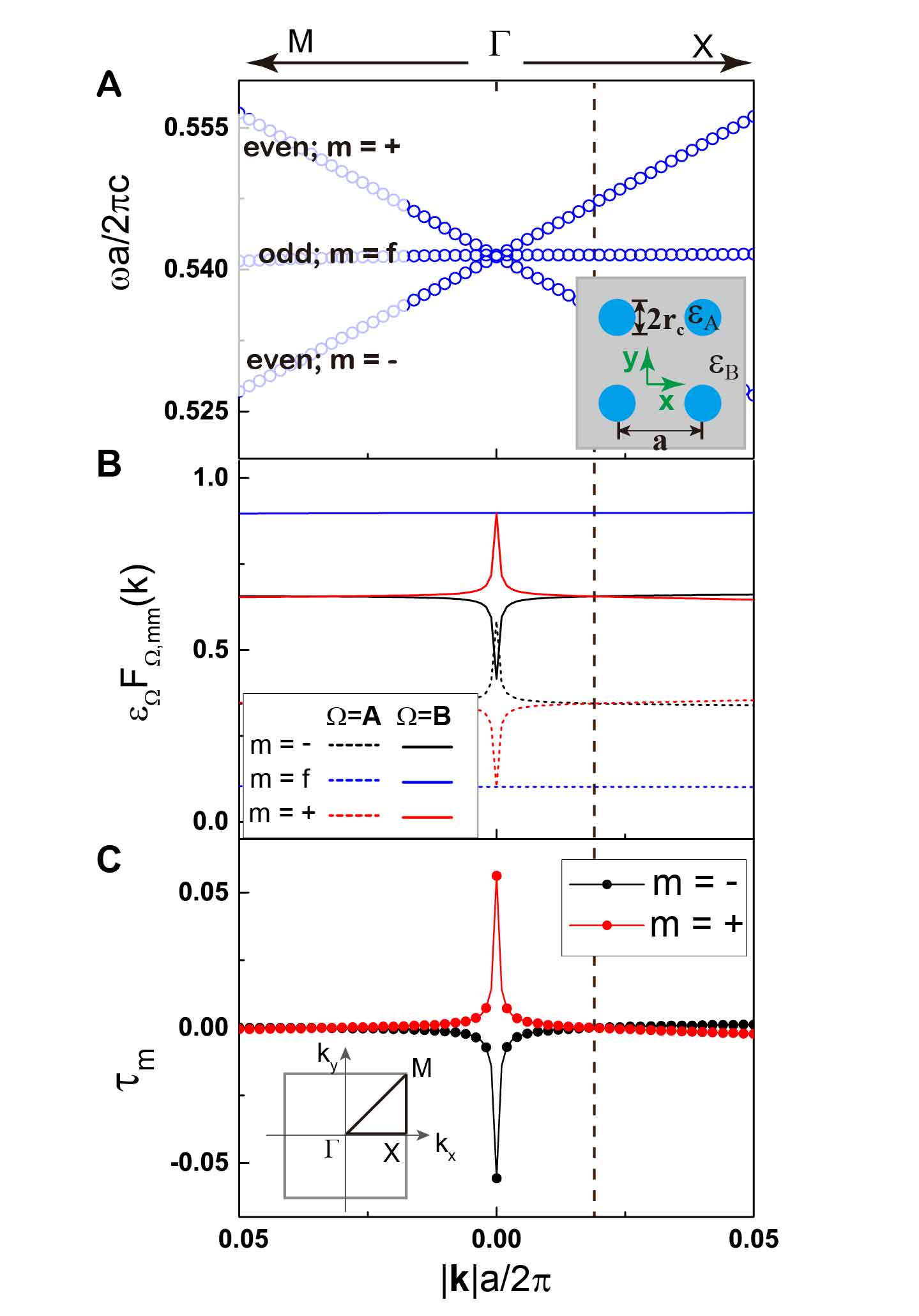}
	\caption{Properties of the Dirac-like cone. (A) Band dispersions of Hermitian PC ($\gamma=0$) calculated using COMSOL. (B) Eigenmode profiles of the upper bands  $ m = + $ (even mode), flat band $ m=f $ (odd mode), and lower band $ m=- $ (even mode). (C) The average non-Hermiticity $ \tau_m $ when the gain-to-loss ratio $ \ell_r=-0.15235 $. The vertical dashed line denotes the intersection: $ F_{\Omega,++} = F_{\Omega,--} $. The inset in (A) shows a schematic of the 2D PC, constructed with cylinders (the blue areas) and the grey area denotes air. The inset in (C) represents the first Brillouin zone of the square lattice. The parameters used are $ r_c=0.1999a $, $ \epsilon_A=12.5+i\gamma $, and $ \epsilon_B=1+i\gamma \ell_r $.
	}
	\label{fig1}
\end{figure}

As shown in the inset of Figure \ref{fig1}A, we consider a two-dimensional (2D) PC with rods arranged in a square lattice in the $ x-y $ plane.  Within the unit cell with a lattice constant $ a $, the rod (domain A, the blue region) having a radius $ r_c $, and relative permittivity $ {\varepsilon _A} = {\varepsilon _{cr}} + i\gamma  $ is embedded in a background medium (domain B, the grey region) with relative permittivity \({\varepsilon _B} = {\varepsilon _{br}} + i{\ell _r}\gamma \). We set $ \varepsilon_{br}=1 $ in the calculation and the relative permeability $ \mu $ of both media as 1.0. The positive (negative) sign of $ \gamma $ indicates that the rod consists of a lossy (active) medium.
A positive sign of $ \ell_r $ indicates that the rod and background are either both lossy or both active, while a negative sign means that one is lossy and the other is active.

We study the PC in the transverse-magnetic (TM) polarization, where the electric field $ \mathbf{E} =E \hat z$ is normal to the $ x-y $ plane, and the electromagnetic waves propagate in the plane. As shown in Figure \ref{fig1}A, in the Hermitian limit ($ \gamma=0 $), this system has a Dirac-like cone dispersion \cite{huang_dirac_2011} at the Brillouin zone center induced by accidental degeneracy. Compared with the Dirac cone of graphene system at the Brillouin zone corner, there is an additional flat band in the Dirac-like cone.
It has been experimentally confirmed that the radiation existing in an open system can spawn rings of EPs in the wave vector space $ \mathbf{k}=(k_x,k_y) $ out of Dirac-like cones \cite{zhen_spawning_2015}; but imaginary parts of the eigenfrequencies are negative because of radiation loss. Now we will see whether we can tune the bands outside the ring to become real spectra by adding gain into the system. 
In principle, the eigenfrequency and eigenfunction of Bloch states for this non-Hermitian 2D PC can be obtained by numerically solving the Helmholtz equation \cite{sakoda_optical_2005}. However, to study the band structure of the non-Hermitian PC analytically, we construct a model Hamiltonian using Hermitian system’s Bloch states obtained at a fixed value of \textbf{k} as the bases, and obtain a generalized eigenvalue problem as (see Supplementary Note 3 for details) \cite{ding_coalescence_2015},
\begin{equation}\label{eq3}
{H_2}{p_{{\bf{k}}n}} = {\left( {{\omega _{{\bf{k}}n}}/c} \right)^2}{H_1} \cdot {p_{{\bf{k}}n}},
\end{equation}
where $c$ is the speed of light, and $\omega_{\bf{k}n}$ and $ {p_{{\bf{k}}n}} $ are the eigenfrequencies and eigenvectors of the non-Hermitian system ($\gamma \ne 0$). The matrices in Eq. \eqref{eq3} are
$	{\left( {{H_1}} \right)_{mm'}}{\rm{ = }}{\delta _{mm'}} + i\gamma \left( {{F_{A,mm'}} + {\ell _r}{F_{B,mm'}}} \right)$, and ${\left( {{H_2}} \right)_{mm'}} = {\delta _{mm'}}{( {\omega _{km}^{( 0)}/c})^2}$. $\omega _{km}^{( 0)}$ is the eigenfrequency of the Hermitian PC. To characterize the eigenmode profiles, we introduce a quantity 
\begin{equation}\label{eq6}
{F_{\Omega ,mm'}} = \int\limits_\Omega  {{{\rm{d}}^2}ru_{{\bf{k}}m}^{\left( 0 \right)*}\left( {\bf{r}} \right)u_{{\bf{k}}m'}^{\left( 0 \right)}\left( {\bf{r}} \right)} ,
\end{equation}
where 
$ \Omega $ denotes the rod ($ |\mathbf{r}|<r_c $) or air ($ |\mathbf{r}|>r_c $) domain. $ {F_{\Omega ,mm'}} $ ($ m\ne m' $ ) expresses the overlapping between two different eigenmodes in the domain $ \Omega $ and $ {F_{\Omega ,mm}} $expresses the amplitude distribution of the eigenmode $ m $ in the domain $ \Omega  $. 
 $u_{{\bf{k}}m}^{\left( 0 \right)}\left( {\bf{r}} \right)$ are the eigenvectors of the Hermitian PC ($\gamma = 0$) and satisfy the orthonormal relation
\begin{equation}
{\varepsilon _A}{F_{A,mm'}} + {\varepsilon _B}{F_{B,mm'}} = {\delta _{mm'}},
\label{eq8}
\end{equation}
where $ {\varepsilon _A}=\varepsilon_{cr} $ denotes the relative permittivity of the rod domain and $ {\varepsilon _B}=\varepsilon_{br}=1 $ represents the relative permittivity of the air domain. 
For convenience, we can omit the subscript \textbf{k} for simplicity and rewrite Eq. \eqref{eq3} as $H \cdot {p_n} = {W_n}{\kern 1pt} {p_n}$, where \({W_n} = {\left( {{\omega _n}/c} \right)^2}\) and 
$H=H_1^{-1}\cdot H_2$.

From Eq. \eqref{eq6}, we know that the function $ F_{\Omega,mm'} $ involves the spatial integrations of eigenfunctions of Hermitian systems, and hence the symmetry of the eigenmode profile plays an important role here. In our system, we choose the center of the rod as origin.
The even/odd symmetry characters are labeled in Figure \ref{fig1}A \cite{sakoda_optical_2005,sakoda_proof_2012}. Since the two linearly dispersive bands ($ m=\pm $) are even while the flat band ($ m=f $) is an odd mode, the matrix elements $ F_{\Omega,+f} $ and $ F_{\Omega,-f} $ are all zero, implying that there is no coupling between the flat band and the linear bands. Therefore, we can decouple the flat band from the Hamiltonian model, and write the Hamiltonian as a $ 2 \times 2 $ matrix (see Supplementary Note 3 for details),
\begin{equation}\label{eq11}
H = \frac{1}{\beta }\left( {\begin{array}{*{20}{c}}
	{W_ - ^{(0)}\left( {1 + i\gamma {\tau _ - }} \right)}&{ - i\gamma \kappa W_ + ^{(0)}}\\
	{ - i\gamma {\kappa ^*}W_ - ^{(0)}}&{W_ + ^{(0)}\left( {1 + i\gamma {\tau _ + }} \right)}
	\end{array}} \right),
\end{equation} 
where $\beta  = 1 + i\gamma \left( {{\tau _ - } + {\tau _ + }} \right) + {\gamma ^2}\left( {|\kappa {|^2} - {\tau _ - }{\tau _ + }} \right)$. In Eq. \eqref{eq11}, 
\begin{equation}\label{eq12}
{\tau _m} = {F_{A,mm}} + {\ell _r}{F_{B,mm}},\,\,\,\,\,\,\,\,\left( {m =  \pm } \right)
\end{equation}
is a real number, representing the average non-Hermiticity of state $ m $ within the primitive unit cell, and
\begin{equation}\label{eq13}
\kappa  = {F_{A, -  + }} + {\ell _r}{F_{B, -  + }},
\end{equation}
is a complex number, representing an overlapping between the two eigenmodes. According to Eq. \eqref{eq12}, the values of $ \tau_\pm $ depend not only on the distributions of non-Hermiticity (described by $ \ell_r $) in the unit cell but also on the eigenmode profiles (as described by $ F_{\Omega,mm} $). The average non-Hermiticity $ \tau_\pm $ is very important, because it can determine whether we can obtain the real spectra in a non-Hermitian PC (see Supplementary Note 4). To see this, we set the average non-Hermiticity $ \tau_\pm=0 $, and then the Hamiltonian \eqref{eq11} becomes
\begin{equation}\label{eq16}
H = \frac{1}{\beta }\left( {\begin{array}{*{20}{c}}
	{W_ - ^{(0)}}&{ - i\gamma \kappa W_ + ^{(0)}}\\
	{ - i\gamma {\kappa ^*}W_ - ^{(0)}}&{W_ + ^{(0)}}
	\end{array}} \right),
\end{equation} 
where $ \beta  = 1 + {\gamma ^2}|\kappa {|^2} $ is a real number. 

Near the $ \Gamma $ point, we can expand the eigenfrequencies of the two linear bands for a small $ k\equiv|\mathbf{k}| $ into 
\begin{equation}\label{eq15}
W_ \pm ^{\left( 0 \right)} = {\left( {{\omega _d}/c \pm {v_g}k/c} \right)^2}\; \approx {W_d} \pm {C_g}k,
\end{equation}
where ${C_g} = 2{\omega _d}{v_g}/{c^2}$, \({W_d} = {\left( {{\omega _d}/c} \right)^2}\), and $ \omega_d $ is the Dirac-like point frequency. The band slope $ v_g $ can be obtained from numerical results in Figure \ref{fig1}A, where we assume $ v_g $ is a constant along all directions in the wave vector space, which is an excellent approximation of a conical dispersion at $ k=0 $ for a system with $ C_{4v} $ symmetry \cite{sakoda_proof_2012}. The corresponding eigenvalues are
\begin{equation}\label{eq17}
{W_ \pm } = \frac{1}{\beta }\left( {{W_d} \pm \sqrt {{{\left( {{C_g}k} \right)}^2}\left( {1 + {\gamma ^2}|\kappa {|^2}} \right) - {\gamma ^2}|\kappa {|^2}{W_d}^2} } \right).
\end{equation}
EPs appears when the value under the square root is zero and form a ring in the wave vector space with a radius \({k_c} = {k_b}(1 + {{\left| \gamma \kappa  \right|}^{ - 2}})^{-\frac{1}{2}}  \le {k_b}\), where \({k_b} = {\omega _d}/2{v_g}\) is the upper bound of the ring’s radius. This means that the radius of the EP ring in \textbf{k} space cannot exceed $ k_b $ for any value of $ \gamma $. In general, $ |\kappa| $ is a small number, and the radius of the ring can be reduced to \({k_c} \approx {k_b}\left| {\gamma \kappa } \right|\). The eigenvalues are real outside the ring ($ k>k_c $) and form complex conjugate pairs inside the ring ($ k < k_c $). Therefore, by setting the average non-Hermiticity $ \tau_\pm=0 $, we can obtain a real spectrum in a certain region (Supplementary Fig. S1b gives a schematic of the band dispersions).

Before proceeding to the  next subsection, the Hamiltonian \eqref{eq16} deserves more comments. Using the linear operator
\begin{equation}
\eta  = \left( {\begin{array}{*{20}{c}}
	{{W_-^{(0)} }}&0\\
	0&{ - {W_ +^{(0)} }}
	\end{array}} \right),
\end{equation}
the Hamiltonian \eqref{eq16} satisfies
\begin{equation}\label{eq19}
{\eta} H\left( \gamma  \right){\eta}^{-1} = H{\left( \gamma  \right)^\dag } =  H\left( { - \gamma } \right),
\end{equation}
implying the Hamiltonian is pseudo-Hermitian, and the operator $ \eta $ transform the system with non-Hermiticity $ \gamma $ into its complex conjugate-paired system $ -\gamma $. Unlike the traditional PT-symmetric system, the loss-gain distribution of our PC system is not PT-symmetric in real space, $n(x) \ne n^*( - x)$.

\section{Results}
\subsection{Realizing the pseudo-Hermitian condition using a Dirac-like cone}
The previous section shows that by setting the average non-Hermiticity $ \tau_\pm=0 $, the non-Hermitian Hamiltonian will become pseudo-Hermitian and we can obtain the real spectra. For a PT-symmetric PC, $ \tau_\pm=0 $ is automatically satisfied since the distributions of loss and gain within the unit cell are equal (see Supplementary Note 2 for details) \cite{szameit_p_2011}. However, for a non-PT-symmetric PC, achieving $ \tau_\pm=0 $ is not an easy task. Now we will show that for a PC exhibiting a Dirac-like cone in the Brillouin zone center, we can always achieve the pseudo-Hermitian condition $ \tau_\pm=0 $.

The eigenmode $u_{{\bf{k}} \pm }^{\left( 0 \right)}\left( {\bf{r}} \right)$ in the quantity $ F_{\Omega,mm'} $ depends on the structural details (such as the filling ratio $r_c$ and the contrast between $\varepsilon_A$ and $\varepsilon_B$) of the PC and can be computed numerically using COMSOL. Substituting the orthonormal relationship Eq. \eqref{eq8} into the pseudo-Hermitian criteria $ \tau_\pm=0 $, we solve a specific value of the loss-gain ratio $ \ell_r $ 
\begin{equation}\label{eq20}
{\ell _r} = \frac{{ - {F_{A,mm}}}}{{{F_{B,mm}}}} = \frac{{ - {F_{A,mm}}{\varepsilon _B}}}{{1 - {F_{A,mm}}{\varepsilon _A}}},
\end{equation}
for both $ m=+ $ and $ m=- $. Equation \eqref{eq20} shows that to determine the value of $ \ell_r $, we must tune the system parameters so that the system satisfies $ F_{\Omega,++}=F_{\Omega,--}  $.

The band dispersions near the Dirac-like cone are shown in Figure \ref{fig1}A, and the quantities $ \varepsilon_\Omega F_{\Omega,mm} $ of the three bands are plotted in Figure \ref{fig1}B. We can see that there is an intersection ($ F_{\Omega,++}=F_{\Omega,--}  $) at ${k_x}a/2\pi  = 0.019$ (marked by the dashed line), and then Eq. \eqref{eq20} can be used to determine the gain-to-loss ratio, which is found to be $ \ell_r=-0.15235 $. Physically, the existence of such an intersection is guaranteed by the band inversion in the Dirac-like cone \cite{huang_dirac_2011}. The condition $ F_{\Omega,++}=F_{\Omega,--}  $ can always be achieved near the Dirac-like cone (see Supplementary Note 6 and Supplementary Figure S1 for details), and therefore, $ \tau_\pm=0 $ is realized at ${k_x}a/2\pi  = 0.019$ as shown in Figure \ref{fig1}C. Note that the quantity $F_{\Omega,mm}$ should be a function of \textbf{k}, but near the Dirac-like cone (except the $\Gamma$ point), this quantity is almost a constant (as shown in Figure \ref{fig1}B), because the eigenmodes of the same point group are almost unchanged when the change of \textbf{k} is small. We can therefore assume the values $\tau_m$, and $\kappa$ are independent of \textbf{k} in the considered Brillouin zone region. As shown in Figure \ref{fig1}C, by tunning $ \ell_r=-0.15235 $, we can obtain $ {\tau _ \pm } \approx 0 $ near the Dirac-like cone (except very close to the $ \Gamma $ point, where $\tau_- \approx -\tau_+$). We can demonstrate that the pseudo-Hermitian condition at the $ \Gamma $ point  is $ \tau_-=-\tau_+ $, which is less demanding than other points, because the symmetry at the $ \Gamma $ point is higher than that at other \textbf{k} points (see Supplementary Note 5 for details). So the pseudo-Hermitian criterion is approximately fulfilled for all the \textbf{k} points near the Dirac-like cone.

\begin{figure}[htpb]
	\centering
	\includegraphics[width=\linewidth]{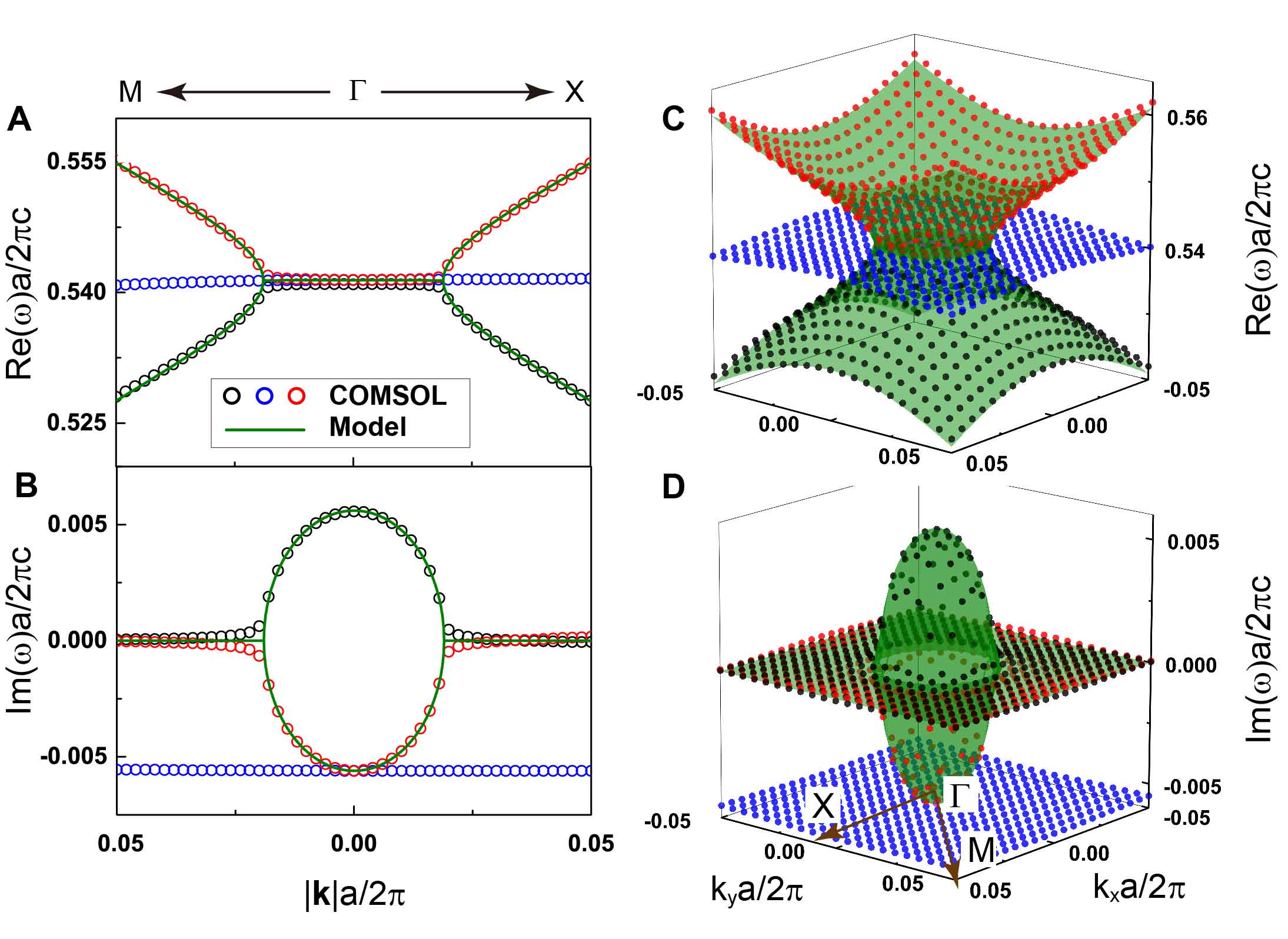}
	\caption{Band dispersions of the pseudo-Hermitian bands. The real parts (A, C) and imaginary parts(B, D) of the complex eigenfrequencies along the $ M-\Gamma-X $ direction and in the 2D Bloch \textbf{k} space. The open circles and solid dots are calculated using COMSOL, while the green solid lines and green surfaces are calculated using the analytical model ($ v_g=0.295c $ and $|\kappa|^2=0.00318$). The bands outside the ring are entirely real. The parameters of the 2D PC are $ r_c=0.1999a $, $ \epsilon_A=12.5+i\gamma $, $ \epsilon_B=1+i\gamma \ell_r $, $\gamma=+0.367$, and $\ell_r=-0.15235$.}
	\label{fig2}
\end{figure}

Now we can calculate the band structures of a PC with loss-gain ratio $\ell_r=-0.15235$ to verify the validity of the pseudo-Hermitian Hamiltonian shown in Eq. \eqref{eq16}. We choose a non-Hermitian strength $ \gamma=+0.367 $ and plot the complex band structure along the $ M-\Gamma-X $ direction in Figures \ref{fig2}A and \ref{fig2}B. The analytical results are shown by the solid green lines and the numerically results calculated using COMSOL by open circles. The validity of our non-Hermitian Hamiltonian model is demonstrated by the good agreement between the two sets of results. We can see that EPs appear in both the $ \Gamma-X $ and $ \Gamma-M $ directions. To show that these EPs form a ring in \textbf{k} space, we plot the three-dimensional complex band structure in Figures \ref{fig2}C and \ref{fig2}D using Eq. \eqref{eq17} and COMSOL, as shown by green surfaces (analytical results) and filled dots (numerical results), respectively. The analytical model \eqref{eq17} predicts that $ k=k_c=0.019(2\pi/a) $ form a ring of EPs, which is verified by the COMSOL results. The eigenfrequecies inside the ring form complex-conjugate pairs, and outside the ring, we obtain the entirely real spectra. 

However, entirely real spectra does not mean loss-gain compensation. To see this, we compare the system with non-Hermiticity $ \gamma_+=+0.367 $ with its complex conjugate-paired system $ \gamma_-=-0.367 $. If we replace $ \gamma_+ $ by $ \gamma_- $, the eigenfrequencies described by Eq. \eqref{eq17} are the same but the Hamiltonians \eqref{eq16} are different as described by Eq. \eqref{eq19}. In other words, these two systems ($ \gamma_+ $ and $ \gamma_- $) possess the same band dispersion, but the eigenfunctions are different (see Supplementary Note 7 for details) and hence they can display different scattering behaviors, as we will show later.

\subsection{Real spectra and complex conjugate media}
The optical properties of a Hermitian PC carrying a Dirac-like cone in the Brillouin zone center can be described using an effective refractive index medium in which the effective permittivity and permeability are simultaneously zero at the Dirac-like point frequency. We have shown in the previous subsection that by setting the average non-Hermiticity as zero, the spectra are real outside the ring of the EPs. It is interesting to see what type of effective medium corresponds to the non-PT-symmetric PC with real spectra. In this section, we will establish the EMT of non-Hermitian PC for describing the physics near the $ \Gamma $ point and study the related scattering properties.

\begin{figure}[htpb]
	\includegraphics[width=\linewidth]{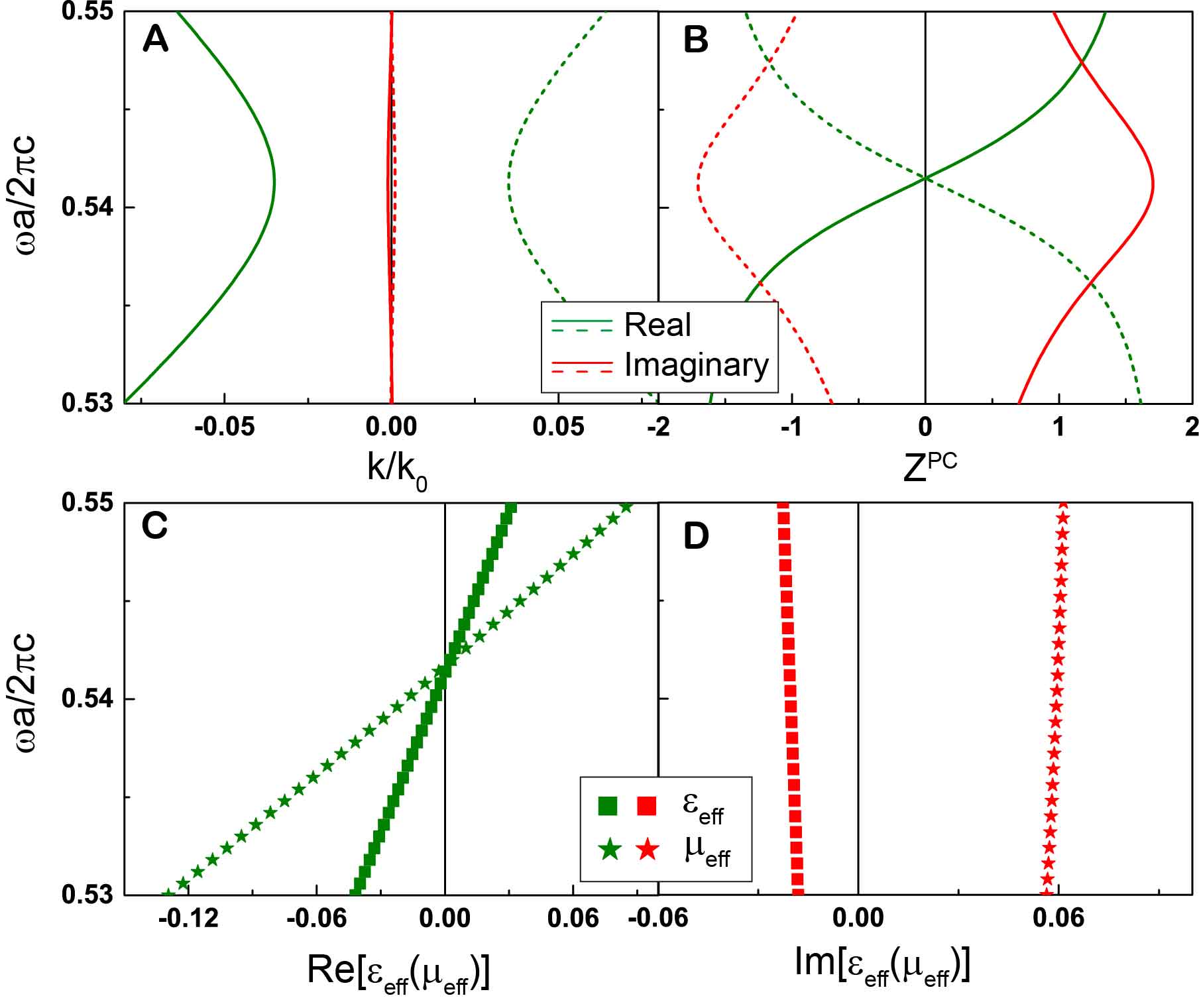}
	\caption{Effective parameters obtained from the real spectra shown in Figure \ref{fig2}. (A) The complex-valued $\textbf{k}(\omega)/k_0$ band and (B)the averaged field ratio $Z^{\rm{PC}}$ calculated numerically from COMSOL. The bands with positive and negative wave vectors are represented by dashed and solid lines respectively. The real (green) and imaginary parts (red) of the effective parameters $ \varepsilon_{\rm{e}} $ and $ \mu_{\rm{e}} $ are plotted in squares and in stars, respectively in (C) and (D).
	}
	\label{fig3}
\end{figure}

We calculate the effective parameters using a boundary field averaging method \cite{andryieuski_bloch-mode_2012}. Assuming that the wave $ \textbf{k}=k \hat x $ propagates along the $ x $ direction, we define \(Z = {E_z}/{H_y} =  - \omega \mu /k =  - k/(\omega \varepsilon )\) for a plane wave traveling in a homogeneous medium to express the ratio between the electric fields and the magnetic fields. For the inhomogeneous PC system, the average field ratio is defined as
\begin{equation}\label{eq21}
{Z^{\rm PC}} = \frac{{\int_{\rm{I}} {E_z^{\rm PC}dy} }}{{\int_{\rm{I}} {H_y^{\rm PC}dy} }},
\end{equation}
where $ E^{\rm PC} $ and $ H^{\rm PC} $ are obtained from the eigenfields at the incident boundary I (the boundary of the unit cell along y direction). When we calculate effective parameters of the non-Hermitian PC as functions of frequency, the frequencies should take real-valued numbers. Because in actual experiments, the incident light comes with a real frequency. Therefore, within the PC, the eigenfields $ E^{\rm PC} $ and $ H^{ \rm PC} $ are obtained by solving the “complex-valued $ \mathbf{k}(\omega) $ vs. real-valued $ \omega $” band structures \cite{davanco_complex_2007}. In Figure \ref{fig3}A, we plot the “complex-valued $ \mathbf{k}(\omega) $” band structure of the non-Hermitian 2D PC calculated using COMSOL. We note that the imaginary parts of the bands (red lines) are almost zero, and for real bands, the “complex-valued $ \mathbf{k}(\omega) $ ” band structure should agree well with the “complex-valued $ \omega(\mathbf{k}) $ vs. real-valued \textbf{k}” band structure (see Supplementary Figure S7 for details). In Figure \ref{fig3}B, we plot $ Z^{\rm PC} $ defined in Eq. \eqref{eq21} for the the bands with positive (dashed lines) and negative (solid lines) wave vectors  respectively in Figure \ref{fig3}A. The effective permittivity and permeability can be calculated as follows:
\begin{equation}\label{eq22}
{\varepsilon _{{\rm{e}}}} = \frac{{ - k}}{{\omega {\varepsilon _0}{Z^{\rm{PC}}}}},\qquad {\mu _{{\rm{e}}}} = \frac{{ - k}}{{{\mu _0}\omega }}{Z^{\rm{PC}}}.
\end{equation}
We plot the real and imaginary parts of $ {\varepsilon _{{\rm{e}}}}({\mu _{{\rm{e}}}}) $ by squares (stars) in Figs. \ref{fig3}C and \ref{fig3}D, respectively. Note that the effective parameters obtained from the two bands (\textbf{k} and -\textbf{k}) are the same.

It is known that for the Hermitian case, the effective parameters of the PC are real and approach zero near the Dirac-like point frequency, indicating that the PC can be treated as a zero refractive index medium \cite{huang_dirac_2011}. When we introduce loss and gain into the PC, the effective parameters  $ \varepsilon_{\rm{e}} $ and $ \mu_{\rm{e}} $ obtain imaginary parts as shown in Figure \ref{fig3}D. However, we note that the effective refractive index \({n_{{\rm{e}}}}^2 = {\varepsilon _{{\rm{e}}}}{\mu _{{\rm{e}}}} = {(k/{k_0})^2}\) is a real number, where $ k_0 $ is the wave vector in air. This automatically indicates the inhomogeneous PCs behaves like the homogeneous CCM whose refractive index is real but whose permittivity and permeability are complex numbers. The real effective refractive index does not necessarily imply a simple loss-gain compensation, because  $ \varepsilon_{\rm{e}} $ and $ \mu_{\rm{e}} $ have imaginary parts. And it can be proved that \(\varepsilon _{{\rm{e}}}^{({\gamma _ + })} = \varepsilon _{{\rm{e}}}^{({\gamma _ - })*}\) and \(\mu _{{\rm{e}}}^{({\gamma _ + })} = \mu _{{\rm{e}}}^{({\gamma _ - })*}\) (see Supplementary Note 8 for details). 



\begin{figure}[htpb]
	\centering
	\includegraphics[width=\linewidth]{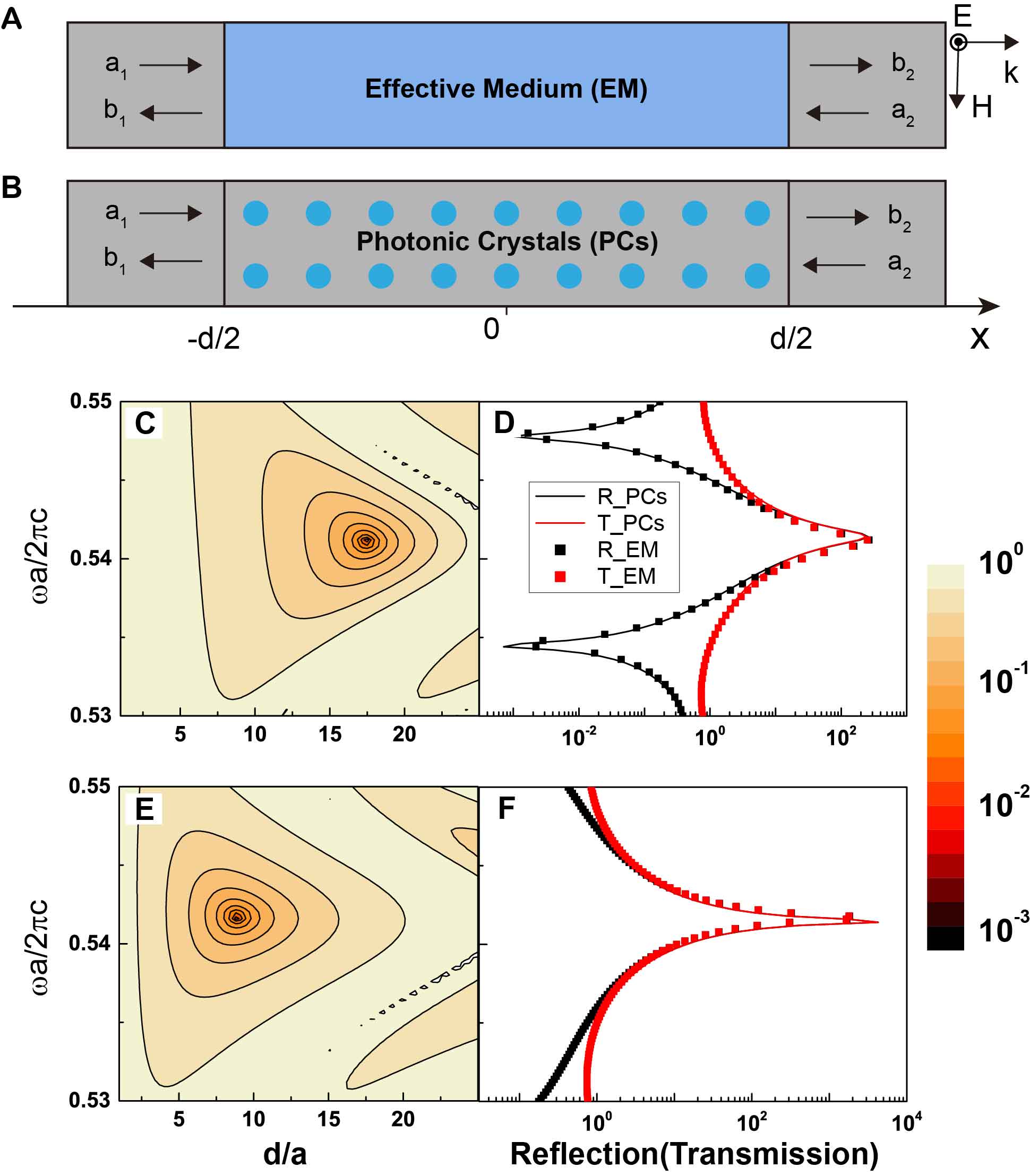}
	\caption{Scattering properties of the CCM with effective parameters shown Figure \ref{fig3}. The schematics of wave scattering in a (A) EM slab and (B) PCs slab. The contour plot of the value $ 1/\rm{Max}|\lambda_\pm| $ in the ($ \omega $, $ d $) plane for the EM slab with parameters (C) $ \varepsilon_{\rm{e}}^{(\gamma_+)} $,  $ \mu_{\rm{e}}^{(\gamma_+)} $ and (E) $ \varepsilon_{\rm{e}}^{(\gamma_-)} $, $ \mu_{\rm{e}}^{(\gamma_-)} $. The black and red solid lines represent respectively the reflection $R$ and transmission $T$ for PCs slab, while the black and red squares represent EM slab. The non-Hermiticity strength of PCs are (D) $ \gamma_+=+0.367 $ and (F) $ \gamma_-=-0.367 $. The slab thicknesses are $ d/a=17 $ in (b) and $ d/a=9 $ in (d).}
	\label{fig4}
\end{figure}

The complex conjugate-paired systems ($ \gamma_+ $ and $ \gamma_- $) have dramatically different effects on the scattering properties of electromagnetic waves. To demonstrate this, we consider a slab with thickness $ d $ formed by the homogeneous effective media (EM) with the parameters $ \varepsilon_{\rm{e}} $ and $ \mu_{\rm{e}} $, as shown in Figure \ref{fig4}A. We assume that plane waves propagate along the $ x $ direction with an electric field polarized along the $ z $ direction and that the EM slab (the blue region) is embedded in air (the grey region). The electric field in the air can be written as $E(x) = {a_1}\exp \left( {i{k_0}x} \right) + {b_1}\exp \left( { - i{k_0}x} \right)$ for $ x<-d/2 $, and as $E(x) = {b_2}\exp \left( {i{k_0}x} \right) + {a_2}\exp \left( { - i{k_0}x} \right)$ for $ x>d/2 $. The amplitudes of incoming-propagating waves $ (a_1, a_2)^T $ and outcoming-propagating waves $ (b_2, b_1)^T $ are related by a scattering matrix \cite{bai_simultaneous_2016}
\begin{equation}\label{eq23}
S = \left( {\begin{array}{*{20}{l}}
	t&r\\
	r&t
	\end{array}} \right)= \frac{1}{{{M_{22}}}}\left( {\begin{array}{*{20}{c}}
	1&{{M_{12}}}\\
	{{M_{12}}}&1
	\end{array}} \right),
\end{equation}
where $ r $ and $ t $ represent the transmission and reflection coefficients of the slab for normal incident plane waves. The elements in S are
\begin{equation}\label{eq25}
{M_{12}} = i\frac{{{n_{{\rm{e}}}}^2 - {\mu _{{\rm{e}}}}^2}}{{2{n_{{\rm{e}}}}{\mu _{{\rm{e}}}}}}\sin \left( {kd} \right),
\end{equation}
\begin{equation}\label{eq26}
{M_{22}} = {e^{\left( {id{k_0}} \right)}}\left[ {\cos \left( {kd} \right) - i\frac{{{n_{{\rm{e}}}}^2 + {\mu _{{\rm{e}}}}^2}}{{2{n_{{\rm{e}}}}{\mu _{{\rm{e}}}}}}\sin \left( {kd} \right)} \right],
\end{equation}
where $ k=k_0 n_{\rm {e}} $ is the wave vector in the slab. The corresponding eigenvalues of the S matrix are
\begin{equation}\label{eq27}
{\lambda _ \pm } = \frac{1}{{{M_{22}}}}\left[ {1 \pm {M_{12}}} \right].
\end{equation}
These eigenvalues can be used to determine the S matrix poles ($1/{\rm{Max}}|{\lambda _ \pm }| \to 0$) where lasing occurs. For a slab composed of CCM [${\mathop{\rm Im}\nolimits} ({n_{{\rm{e}}}}) = {\mathop{\rm Im}\nolimits} (k/{k_0}) = 0$], $ kd $ is a real number. For the S matrix to have poles, we require $ M_{22}=0 $, and therefore we obtain
\begin{equation}\label{eq28}
\frac{{{n_{{\rm{e}}}}^2 + {\mu _{{\rm{e}}}}^2}}{{{n_{{\rm{e}}}}{\mu _{{\rm{e}}}}}} =  - 2i\cot \left( {kd} \right).
\end{equation}
To satisfy Eq. \eqref{eq28}, $ \mu_{{\rm{e}}} $ must be an imaginary number. Figures \ref{fig3}C and \ref{fig3}D show the effective parameters of PCs with non-Hermiticity $ \gamma_+=+0.367 $. We find ${\mathop{\rm Re}\nolimits} [\mu _{{\rm{e}}}^{\left( {{\gamma _ + }} \right)}] = 0$ at $\omega a/2\pi c = 0.5416$, and we can solve Eq. \eqref{eq28} to obtain the slab thickness $ d/a=17.4 $. For PCs with non-Hermiticity $ \gamma_-=-0.367 $, the effective parameters are \(\varepsilon _{{\rm{e}}}^{({\gamma _ + })} = \varepsilon _{{\rm{e}}}^{({\gamma _ - })*}\) and \(\mu _{{\rm{e}}}^{({\gamma _ + })} = \mu _{{\rm{e}}}^{({\gamma _ - })*}\), and we can solve the slab thickness $ d/a=8.9 $.

Figures \ref{fig4}C and \ref{fig4}E show the contour plots of $ 1/\rm{Max}|\lambda_\pm| $ in the ($ \omega a/2 \pi c, d/a $) plane for EM slab with $(\varepsilon _{{\rm{e}}}^{\left( {{\gamma _ + }} \right)},{\kern 1pt} {\kern 1pt} {\kern 1pt} {\kern 1pt} \mu _{{\rm{e}}}^{\left( {{\gamma _ + }} \right)})$ and $(\varepsilon _{{\rm{e}}}^{\left( {{\gamma _ - }} \right)},{\kern 1pt} {\kern 1pt} {\kern 1pt} {\kern 1pt} \mu _{{\rm{e}}}^{\left( {{\gamma _ - }} \right)})$ respectively. To verify the scattering poles ($ 1/\rm{Max}|\lambda_\pm| \to 0$), in Figure \ref{fig4}B, we plot the transmission $ T=|t|^2 $ and reflection $ R=|r|^2 $ of slabs formed by homogeneous EM by squares and PCs by solid lines. The schematic of the one-dimensional scattering problem of slab formed by PCs is plotted in Figure \ref{fig4}B. We set the thickness of the slab $ d/a = 17$, because the thickness of the slab formed by PCs must be an integer. The good agreement between the EM and PCs results shows that the effective parameters shown in Figs. \ref{fig3}C and \ref{fig3}D essentially reproduce the physics. We can see that the system indeed has the transmission and reflection singularities at $ \omega a/2 \pi c=0.5416 $ in Figure \ref{fig4}D. This indicates that we can observe the poles of the S matrix from the transport configuration. Similarly, the pole in Figure \ref{fig4}E at $ d/a=8.9 $ and   $ \omega a/2 \pi c=0.5416 $ for $(\varepsilon _{{\rm{e}}}^{\left( {{\gamma _ - }} \right)},{\kern 1pt} {\kern 1pt} {\kern 1pt} {\kern 1pt} \mu _{{\rm{e}}}^{\left( {{\gamma _ - }} \right)})$ is also verified by the transmission and reflection singularities in Figure \ref{fig4}F, where $ d/a=9 $. When we change the sign of $ \gamma $, the real bands of the PCs remains almost the same but the non-Hermitian PCs do not have time reversal symmetry. Although the refractive index satisfies \(n_{{\rm{e}}}^{\left( {{\gamma _ + }} \right)} \approx n_{{\rm{e}}}^{\left( {{\gamma _ - }} \right)}\), the scattering phenomena change drastically because the effective permeability (or impedance) has imaginary parts, and \(\mu _{{\rm{e}}}^{({\gamma _ + })} \ne \mu _{{\rm{e}}}^{({\gamma _ - })}\). This illustrates that the S matrix poles depend not only on the refractive index but also on the permeability as shown in Eq. \eqref{eq28}. The sum of the transmission and reflection is not equal to unity, although the slab’s effective refractive index is real. A real refractive index does not necessarily imply energy compensation. The focusing effect of our PCs as a lens is also different for these complex paired systems (see Supplementary Note 9 for details).

\section{Discussion}
In this work, we find that a non-PT-symmetric PC can have real spectra when the average non-Hermiticity within the unit cell is zero. We showed that the pseudo-Hermitian condition ($ \tau_\pm=0 $) can always be fulfilled in two-component PCs carrying Dirac-like cones with bands arising from monopolar and dipolar resonances. These non-Hermitian PCs carrying Dirac-like cones can be used to realize CCM near the Dirac-like point frequency, where a real spectrum guarantees a real refractive index and the non-Hermiticity guarantees complex permittivity and permeability.

\section{acknowledgement}
This work is supported by the Research Grants Council, University Grants Committee, Hong Kong, through grants No. AoE/P-02/12, N\_HKUST608/17, 16303119 and C6013-18G and by the National Natural Science Foundation of China (Grant Nos. 11761161002, 61775243). K.D. acknowledges funding from the Gordon and Betty Moore Foundation. We would like to thank Prof. Zhao Qing Zhang  and Dr. Ruoyang Zhang for helpful discussions.

\bibliographystyle{unsrt}
\bibliography{non-Hermitian}
\end{document}